\documentclass[twocolumn,showpacs,preprintnumbers,amsmath,amssymb]{revtex4}
\usepackage{graphicx}% Include figure files
\usepackage{dcolumn}% Align table columns on decimal point
\usepackage{bm}% bold math
\usepackage{here}

\begin{document}
\draft

\title{Vibrational modes and spectrum of oscillators on a scale-free network}

\author{Kazumoto Iguchi\cite{byline} and Hiroaki Yamada\cite{byline2}}

\address{{\it 70-3 Shinhari, Hari, Anan, Tokushima 774-0003, Japan\/}\cite{byline}}
\address{{\it 5-7-14 Aoyama, Niigata 950-2002, Japan\cite{byline2}\/}}

\date{\today}

\begin{abstract}
We study vibrational modes and spectrum of a model system of atoms and springs
on a scale-free network in order to understand the nature 
of excitations with many degrees of freedom on the scale-free network.
We assume that the atoms and springs are distributed as nodes and links
of a scale-free network, assigning the mass $M_{i}$ and the specific oscillation 
frequency $\omega_{i}$ of the $i$-th atom and the spring constant $K_{ij}$ 
between the $i$-th and $j$-th atoms. 
\end{abstract}

%\begin{multicols}{2}
%\narrowtext
\maketitle

\section{Introduction}
Recently there has been a notable progress in the study of the so-called 
scale-free network (SFN)\cite{FFF,Barabasi,AB,KRL,KRR,DMS,DM,KK,BE,BCK}.
In the point of view of network theory, 
the random network theory (RN) was
first invented by  Erd\"{o}s and R\'{e}nyi\cite{ER}
and has been applied to many areas of sciences from physics such as
Anderson localization\cite{Ander},
percolation\cite{Ziman}, 
and
energy landscape\cite{WMW,SDS}
to biology such as the Kauffman's NK model\cite{Kauf}.
Recently it was generalized to the small world network (SWN) models
\cite{Klein,WS,NMWS,CNSW,NSW,BA,ASBS,LCS,Strog} 
and has been applied to
physical systems such as
percolation\cite{Klein,WS,NMWS,CNSW,NSW,BA,ASBS,LCS,Strog}, 
arrays of coupled oscillators with synchronization known as the Kuramoto model\cite{Kura}, 
arrays of coupled lasers\cite{LE},
arrays of the Josephson junctions\cite{WCS}, 
electric circuit\cite{CJS}, 
traffic transportation\cite{BMR,BCFMRR}, and
protein folding\cite{SAB}.
And very recently, the SFN was discovered by studying the network
geometry of the internet\cite{FFF,Barabasi,AB,LG,HPPL,AJB1,AJB2}.
Albert, Jeong and Barab\'{a}si\cite{Barabasi,AB,LG,HPPL,AJB1,AJB2} 
opened up an area for studying
very complex and growing network systems such as 
internet\cite{Klein,LG,HPPL,AJB1,AJB2,CEBH}, 
biological evolution\cite{SK},
metabolic reaction\cite{AHB},
epidemic disease\cite{PV},
human sexual relationship\cite{LEASA}, and 
economy\cite{Arthur}.
These are nicely summarized in the reviews by Barab\'{a}si\cite{Barabasi}.

%\section{AB-model}
The nature of these SFNs is characterized by
the power-law behavior of the distribution function
for the number of nodes with $k$ links such as 
$P(k) \propto k^{-\gamma}$ where $\gamma \approx 1-4$.
In order to show the power-law distribution of the SFN,
Albert and Barab\'{a}si first proposed a very simple model called
the Albert-Barab\'{a}si (AB)'s SFN model\cite{Barabasi,AB,LG,HPPL,AJB1,AJB2}.
This system is constructed by the following process:
Initially we put $m_{0}$ nodes as seeds for the system.
Every time when a new node is added, $m$ new links are
distributed from the node to the existed nodes in the system
with a preferential attachment probability
$\Pi_{i}(k_{i})=k_{i}/\sum_{i=1}^{N-1}k_{i}$, 
where $k_{i}$ is the number of links at the $i$-th node
and we have assumed $m \le m_{0}$. 
The development of this model is described 
by a continuum model
$\frac{dk_{i}}{d\tau} = m\Pi_{i}(k_{i}) = \frac{m k_{i}}{2\tau}$. 
Then at time $\tau$ the system consists of $N(\tau)$ nodes and the $L(\tau)$ links 
with 
$L(\tau)= \frac{1}{2}\sum_{i=1}^{N(\tau)}k_{i}$.
This model exhibits $\gamma = 3$ for the power-law.
Thus, it has been concluded that
the essential points of why a network grows to a SFN
are attributed to the growth of the system and the  preferential attachment 
of new nodes to old nodes existed already in the network.

%\section{motivation}
However, although time evolution of the SFN has been 
intensively studied regarding nodes and links as metaphysical objects 
such as agents and relationships in an area of science,
it seems that very few physical models putting
real meaning on nodes and links in the SFNs have been  
studied in order to investigate excitations such as vibrations, 
phonons, and electrons, except diffusion\cite{NR} and spins\cite{Kraw} on the SFN
and excitations in the RN\cite{ER,Ander,Ziman,WMW,SDS,Kauf} and 
SWN\cite{Klein,WS,NMWS,CNSW,NSW,BA,ASBS,LCS,Strog}.
So, we explore to study, as a prototype model, vibrational modes and spectrum
of a system of atoms coupled by springs
where the atoms and springs are located regarding 
as nodes and links of a SFN.

\section{vibrational model}
Let us introduce our vibrational model.
We first adopt the AB's SFN model for the construction of a SFN
and we regard nodes and links in the SFN as atoms and springs in our physical model.
Assuming that $q_{i}$ and $\omega_{i}$ are the displacement and 
the specific frequency of the $i$-th atom of mass $M_{i}$, respectively,
we can define the hamiltonian of the system:
$$H = \sum_{i=1}^{N(\tau)} \left(\frac{M_{i}}{2}\dot{q}_{i}^{2}
+ \frac{M_{i}\omega_{i}^{2}}{2} q_{i}^{2} \right)
+ \sum_{i,j=1(i\ne j)}^{N(\tau)} \frac{K_{ij}}{2} (q_{i}-q_{j})^{2}, \eqno{(1)}$$
where $\dot{q}_{i}=\frac{d q_{i}}{dt}$ the velocity of the $i$-th atom 
and $K_{ij}$ is the spring constant between 
the $i$-th and $j$-th atoms with $K_{ij} = K_{ji}$.
Although we will concentrate to study only this atom-spring model in this paper, 
the generalizations of this model and the applications to other systems are straightforward.
We expect that the physical nature of this model shares with those of such models and systems.

We now assume that the time interval
$\Delta \tau$ for the development of link addition process
is much larger than that $\Delta t$ of the physical model such that 
$|\Delta \tau| \gg |\Delta t|$.
This guarantees that although the network grows in the course of its development,
as long as the network consists of $N(\tau)$ nodes and $L(\tau)$ links,
the vibrational model can be simultaneously solved.
This means that {\it time evolution the network is adiabatic to the
time motion of the atoms and springs\/}.
By using the Euler-Lagrange equation,
$\frac{d}{dt}(\frac{\partial H}{\partial \dot{q}_{i}})=\frac{\partial H}{\partial q_{i}}$,
we obtain
$$M_{i} (\ddot{q}_{i} + \omega_{i}^{2}q_{i}) = \sum_{j=1}^{N(\tau)}K_{ji}(q_{j}-q_{i}), \eqno{(2)}$$
for $i = 1, \dots , N(\tau)$.
Assuming 
$q_{i}(t) = q_{i}(\omega)e^{-i\omega t}$, Eq.(2) becomes
$$M_{i} (\omega_{i}^{2}-\omega^{2})q_{i} = \sum_{j=1}^{N(\tau)}K_{ji}(q_{j}-q_{i}), \eqno{(3)}$$
for $i = 1, \dots , N(\tau)$.
This is the eigenequation for our system.

%\section{simple assumption}
Let us assume that all springs are identical for the sake of simplicity
such that 
$K_{ij} = K_{0}A_{ij}$,
where $K_{0}$ is the spring constant and 
$A_{ij}$ is the $ij$-th component of the adjacency matrix 
${\bf \hat{A}}$ for the network geometry.
The components of the adjacency matrix are non-negative 
such that $A_{ij} = 0$ or $1$ according to whether or not
a link between the $i$-th and $j$-th nodes exist in the network.
The link number $k_{i}(\tau)$ at the $i$-th atom 
(i.e., the order of the $i$-th node) is given by
$k_{i}(\tau) = \sum_{j=1}^{N(\tau)}A_{ji}$.
From this, the last term in Eq.(3) becomes
$\sum_{j=1}^{N(\tau)}K_{ji}q_{i} =  K_{0} k_{i}(\tau)q_{i}$.
Hence, in this setting, we obtain
$$\Omega_{i} q_{i} =  K_{0}\sum_{j=1}^{N(\tau)} A_{ji}q_{j}, \eqno{(4)}$$
for $i = 1, \dots , N(\tau)$, where 
$$\Omega_{i} \equiv M_{i}(\omega_{i}^{2}-\omega^{2}) + K_{0} k_{i}(\tau).\eqno{(5)}$$

\section{Green's function formalism}
Let us now define the Green's function by
$$\sum_{j=1}^{N(\tau)}
\left[ \Omega_{i}\delta_{ij} -  K_{0}A_{ij} \right] G_{jk} = \delta_{ik}, \eqno{(6)}$$
for $i$, $k = 1, \dots , N(\tau)$, which is represented by
$[{\bf\hat{G}}_{0}^{-1} - K_{0}{\bf\hat{A}}]{\bf\hat{G}}={\bf\hat{1}}$
in the matrix representation
where ${\bf\hat{1}}$ is the  $N(\tau) \times  N(\tau)$ unit matrix
and ${\bf\hat{G}}_{0}$ is the $N(\tau) \times  N(\tau)$ diagonal matrix defined by
${\bf\hat{G}}_{0} = \Omega_{i}^{-1}\delta_{ij}$.
Thus, the Green's function is formally  obtained as
${\bf\hat{G}}^{-1} = {\bf\hat{G}}_{0}^{-1} - K_{0} {\bf\hat{A}}$.
Furthermore, we can derive a series expansion of ${\bf \hat{G}}$ 
in terms of ${\bf \hat{G}}_{0}$ and ${\bf \hat{A}}$ as
$${\bf\hat{G}} = {\bf\hat{G}}_{0} + K_{0}{\bf\hat{G}}_{0}{\bf\hat{A}}{\bf\hat{G}}_{0} 
+ K_{0}^{2}{\bf\hat{G}}_{0}{\bf\hat{A}}{\bf\hat{G}}_{0}{\bf\hat{A}}{\bf\hat{G}}_{0} + \cdots$$
$$= {\bf\hat{G}}_{0} + K_{0}{\bf\hat{G}}_{0}{\bf\hat{A}}{\bf\hat{G}}
= {\bf\hat{G}}_{0} + {\bf\hat{G}}_{0}{\bf\hat{T}}{\bf\hat{G}}_{0},\eqno{(7)}$$
where ${\bf\hat{T}}$ is called the $T$-matrix defined as
$${\bf\hat{T}} = K_{0}{\bf\hat{A}} +
 K_{0}^{2}{\bf\hat{A}}{\bf\hat{G}}_{0}{\bf\hat{A}} + \cdots
= {\bf\hat{T}}_{1} + {\bf\hat{T}}_{2} + \cdots. \eqno{(8)}$$
We can now derive the following:
$$({\bf\hat{T}}_{n})_{ij}
= K_{0}^{n}({\bf\hat{A}}{\bf\hat{G}}_{0}{\bf\hat{A}} \cdots {\bf\hat{G}}_{0}{\bf\hat{A}})_{ij}$$
$$= K_{0}^{n} \sum_{j_{1},j_{2}, \cdots ,j_{n-1}}
\frac{A_{ij_{1}}A_{j_{1}j_{2}}\cdots A_{j_{n-1}j}}
{\Omega_{j_{1}}\Omega_{j_{2}}\cdots \Omega_{j_{n-1}}}\equiv K_{0}^{n}\Gamma_{ij}^{(n)},\eqno{(9)}$$
and since $K_{ij} = K_{ji}$ (i.e., $A_{ij} = A_{ji}$), we find
$$({\bf\hat{T}}_{n})_{ij} = ({\bf\hat{T}}_{n})_{ji}. \eqno{(10)}$$
From Eqs.(8) and (9), we find
$({\bf\hat{T}})_{ij} = K_{0}\Gamma_{ij}^{(1)} + K_{0}^{2}\Gamma_{ij}^{(2)} + \cdots. $

Let us consider the trace of the Green's function.
We now get
$$Tr{\bf\hat{G}} = Tr({\bf\hat{G}}_{0} + {\bf\hat{G}}_{0} {\bf\hat{T}}{\bf\hat{G}}_{0})
= \sum_{i}\left( \frac{1}{\Omega_{i}} + \frac{({\bf\hat{T}})_{ii}}{\Omega_{i}^{2}}\right).\eqno{(11)}$$
We note here that if $M_{i} (\omega_{i}^{2}-\omega^{2}) = 0$
such that
$\Omega_{i} = K_{0} k_{i}(\tau)$, 
then
$({\bf\hat{T}}_{n})_{ij} = K_{0} k_{i}P_{ij}$
with
$k_{i}P_{ij} = k_{j}P_{ji}$,
where
$P_{ij}(\tau)$ means the probability that the walker starts at 
node $i$ at time $t = 0$ and found at node $j$ at time $t =\tau$
in terms of the language of the diffusion theory of Noh and Rieger 
[see Eqs(2)  and (3) in \cite{NR}].
As is well-known, the density of states $\rho(\omega)$ is given by
$$\rho(\omega) = - \frac{1}{\pi}Tr{\bf\hat{G}}(\omega+i\epsilon).\eqno{(12)}$$
Thus, the poles of the Green's function produce the spectrum of the system.

\section{Special limits}
Before going to do the direct calculation for spectrum of the system,
let us consider some limits.
(i) {\it The independent atom limit}.
First, in the case of no springs of $K_{0} = 0$, 
since $Tr({\bf\hat{G}}) = Tr({\bf\hat{G}}_{0})$, 
the poles of the Green's function are given by
$\Omega_{j} \equiv M_{j} (\omega_{j}^{2}-\omega^{2}) = 0$,
which trivially provides the discrete spectrum $\omega = \omega_{j}$
for $j = 1, \dots, N(\tau)$.
This means that the atoms independently vibrate with specific frequencies $\omega_{i}$.

(ii) {\it The AB limit}. 
Second, in the case of very weak spring constant such as $K_{0} \ll 1$,
the the poles of the Green's function are obtained as
$$\omega^{2} = \omega_{j}^{2} +  \frac{K_{0}}{M_{j}}k_{j}(\tau).\eqno{(13)}$$
This means that each atom vibrates with frequency
related to the number of links of the atom.
Since the distribution of the nodes with $k$ links is
given by $P(k) \propto k^{-\gamma}$ in the SFN\cite{Barabasi,AB},
the distribution of the spectrum is given as
$$P(\omega^{2} - \omega_{j}^{2}) \propto (\omega^{2} - \omega_{j}^{2})^{-\gamma}.\eqno{(14)}$$
Hence, this limit shares with the nature of the AB's SFN geometry.
Therefore, we may call this limit the AB limit.

(iii) {\it The localized mode limit}.
Third, let us consider the limit of very small mass ($M_{i} \ll 1$) or
very strong spring ($K_{0} \gg 1$).
In this case, we can ignore the frequency dependence in the eigenequation of Eq.(4)
such as $\Omega_{i} \approx K_{0}k_{i}$,
which then yields
$$k_{i} q_{i} =  \sum_{j=1}^{N(\tau)} A_{ji}q_{j}, \eqno{(15)}$$
for $i = 1, \dots, N(\tau)$.
Since we can rewrite the above equation as
$\sum_{j=1}^{N(\tau)}[k_{i}\delta_{ij} - A_{ij}] q_{j} = 0$,
non-trivial solutions may exist only when the determinant 
$\det[k_{i}\delta_{ij} - A_{ij}]$ 
vanishes.
This is realized when 
$q_{j} = q_{i}$ where $j$ runs the adjacent links around the $i$-th atom. 
In this sense the mode is localized within the adjacent atoms.

\section{calculations of the spectrum}
Let us now calculate the spectrum of the system 
of oscillators in the SFN.
This is carried out by directly diagonalizing Eq.(4).
For the sake of simplicity,
we assume that
$M_{i}=M_{0}=$ const. and $\omega_{i} = \omega_{0} = $ const.
and we adopt the AB-model for generating the SFN geometry.
We have performed the calculations for the systems up to $N = 10^{4}$.

Fig.1 (a) shows the density of states of the system,
where we have calculated for the case of $m = 2$ and $N =10^{4}$ (blue)
and the case of $m = 4$ and $N = 10^{4}$ (red),
respectively, where we have used $m_{0} = 5$.
To obtain the distributions, we have used 
twenty configurations with different random numbers.
The vertical axis means $\rho(\omega)\sqrt{\langle k \rangle_{2}}$,
while the horizontal axis means $\omega^{2}/\sqrt{\langle k \rangle_{2}}$,
where $\langle k \rangle_{2}$ stands for 
the second order average degree of a node [See Eq.(18)].
The shape of the curve is unique such that
there is a peak at $\omega = \omega_{0}$
and the spectral tail exists in the whole range of the spectrum.
This tendency means that there is a scale-free nature
in the spectrum of the vibrational modes in the system.

%%% Figure 1 %%%
\begin{figure}[h]
\includegraphics[scale=0.6]{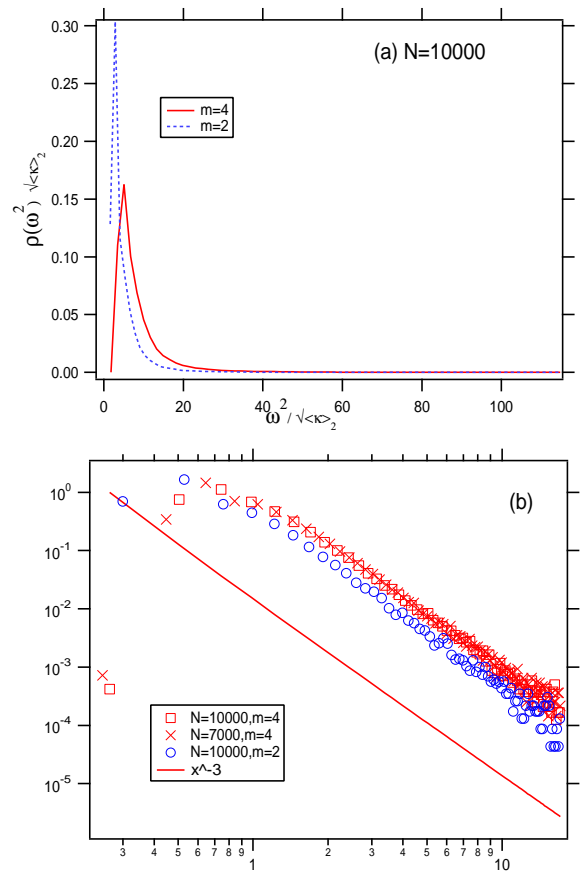}
%\begin{figure}
%\vglue 3.8in
%\hspace{0.1in}
%\special{picture Fig.1 [scaled 0.5]}
%\vglue 0.1in
\caption{
(color online)  
The density of states and its tail behavior
of the system of oscillators in the SFN.
(a)
The density of states is calculated for the cases of the AB-model with $m = 2$ (blue)
and $m = 4$ (red) for $N = 10^{4}$, respectively.
The distributions are obtained as an average over 
twenty configurations with different random numbers.
The vertical axis means the density of states, $\rho(\omega^{2}) \sqrt{\langle k \rangle_{2}}$,
while the horizontal axis means $\omega^{2}$, 
where $\omega$ is the vibrational frequency of the oscillators
and $\langle k \rangle_{2}$ the second order average degree of a node.
(b)
The tail behavior of the density of states 
is shown for the cases of the AB-model with 
$m = 2$ and $N =10^{4}$ (blue circles),
with $m = 4$ and $N =7\times 10^{3}$ (red crosses),
and
with $m = 4$ and $N = 10^{4}$ (red squares), respectively.
The vertical axis means $\log$-plot of the density of states 
while the horizontal axis means $\log$-plot of $\omega^{2}$, 
where $\omega$ is the vibrational frequency of the oscillators.
The red line is a guide for showing $(\omega^{2})^{-3}$.
Here we have assumed that
$\omega_{0} = K_{0} = M_{0} =1$ and $m_{0} = 5$.
}
%\label{fig:1}
\end{figure}
%%%%%

Fig.1 (b) shows the tail behavior of the density of states.
The density of states is shown in a $\log$-$\log$ plot
for the cases of the AB-model 
with $m = 2$ and $N =10^{4}$ (blue circles),
with $m = 4$ and $N =7\times 10^{3}$ (red crosses),
and
with $m = 4$ and $N = 10^{4}$ (red squares), respectively.
The red line is a guide for showing $(\omega^{2})^{-3}$.

From this we find that the tail behavior of Eq.(14) holds valid
for the general cases as well.
Therefore, we can conclude that
our vibrational model shares common nature with
the AB-model of the SFN.
This is contrary to the conclusion previously obtained
from the calculations of the spectrum of the
adjacency matrix ${\bf \hat{A}}$ of the AB-model\cite{FDBV,GKK,DGMS,CLV}.
There, when the network has the tail behavior of
$P(k) \propto k^{-\gamma}$,
the spectral tail for the eigenvalues $\lambda$ of the adjacency matrix
is given by
$\rho(\lambda) \propto \lambda^{-\gamma'}$
where $\gamma' = 2\gamma -1$.
Therefore, since the AB-model has $\gamma = 3$,
we conclude $\gamma' = 5$.
This is different from our result of $\gamma' = 3$.
The main reason for this phenomenon is explained as follows:
In our vibrational model the $\Omega_{i}$ consists
of the degree $k_{i}$ of the node [see Eq.(4)].
Therefore, as the system grows, so does the
magnitude of $\Omega_{i}$.
This can reduce the contributions of the adjacency matrix ${\bf \hat{A}}$
in the higher terms of the perturbation series of Eq.(7).
Hence, the spectral behavior is dominated by the pole of the
unperturbed Green's function ${\bf \hat{G}}_{0}$.
Thus, we are led to the same spectral behavior in the AB-limit.

The physical meaning of the above results can be understood as follows:
The main peak in the density of states is attributed to vibrational modes
with frequency $\omega_{0}$.
These modes are extremely localized within the least connected nodes in the SFN
such that the total number of the localized modes 
provides the height of the peak.
Since the number of modes is nothing but the number of 
degeneracy of the eigenequation, these localized modes are highly degenerate.

On the other hand, there is the power-law tail of 
$\rho(\omega^{2}) \propto (\omega^{2})^{-3}$
as $\omega \rightarrow \infty$.
This means that the larger the frequency of modes
the fewer the number of modes.
In other words, as the frequency is increasing,
the number of modes is decreasing by the power-law.
As the result, there appears only one mode with the maximum 
frequency (i.e., the maximum eigenvalue).
The mode with the maximum frequency is extended over
the entire system of the SFN.
This situation means that in the SFN the lowest frequency modes 
can be very easily excited,
but it is very hard to excite the maximum energy mode.
Thus, the high frequency modes are very hard to 
exist in the system of oscillators coupled in the SFN.
This nature is very different from that of the standard
systems of networks such as RN\cite{ER} and lattices\cite{Ziman} 
that there are a small number of orders of nodes.
This is the most prominent characteristic of our system.

\section{the maximum eigenvalue}
The behavior of the maximum eigenvalue $\lambda_{max}$ 
of the adjacency matrix ${\bf \hat{A}}$ is very important in
the network theory\cite{FFF,Barabasi,FDBV,GKK,DGMS,CLV}. 
In the standard networks such as the random networks\cite{Barabasi,ER},
the maximum eigenvalue $\lambda_{max}$
cannot grow so fast as the network grows\cite{FFF,Barabasi,FDBV,GKK,DGMS,CLV}.
And also, as in solid state physics, networks in most of physical systems 
provide the so-called energy band that is a spectrum with a finite region\cite{Ander,Ziman}.
This is due to the topology of the finite coordination number 
of atoms in the network of the lattice structure\cite{Ander,Ziman}.
So, in order to elucidate the difference between the SFNs and other networks
the growth of the maximum eigenvalue is an important signature.

As was numerically studied by many authors\cite{FFF,Barabasi,AB,FDBV,GKK,DGMS},
the maximum eigenvalue $\lambda_{max}$ of the adjacency matrix 
${\bf \hat{A}}$ in the AB-model is proportional to
$\sqrt{k_{max}}$ such that 
$$\lambda_{max} \propto \sqrt{k_{max}}.  \eqno{(16)}$$
Here $k_{max}$ means the maximum order of nodes in the network
such that 
$k_{max} = \max_{i}\{ k_{i} \}$ 
(We will use this notation for later purposes).
And the numerical studies showed that
$k_{max} \propto \sqrt{N}$.
Therefore, we obtain
$$\lambda_{max} \propto N^{1/4}.           \eqno{(17)}$$

To see whether or not this is true in an arbitrary SFN
and to know how general it is, 
very recently, Chung, Lu and Vu\cite{CLV} have
proved a very general theorem:
\newtheorem{theorem}{Theorem}
\begin{theorem}
Suppose that the distribution of degrees of nodes in a SFN
is represented by $P(k) \propto k^{-\gamma}$.
Denote by $\langle k \rangle_{2}$ the second order average degree of a node.
This is defined by
$$\langle k \rangle_{2} \equiv \frac{\sum_{i=1}^{N(\tau)}k_{i}^{2}}{\sum_{i=1}^{N(\tau)}k_{i}}
= \frac{\langle k^{2} \rangle}{\langle k \rangle},                     \eqno{(18)}$$
where $\langle k^{p} \rangle = \frac{1}{N(\tau)}\sum_{i=1}^{N(\tau)} k_{i}^{p}$ with $p$ integer.
Then, (C1) if the exponent $\gamma > 2.5$, then 
$$const. \langle k \rangle_{2} \le \lambda_{max} \le const.\sqrt{k_{max}}.   \eqno{(19)}$$
(C2) If the exponent $2 < \gamma < 2.5$, then 
$$const. \sqrt{k_{max}} \le \lambda_{max} \le const. \langle k \rangle_{2}.  \eqno{(20)}$$
(C3) And if the exponent $\gamma = 2.5$, then a transition happens.
\end{theorem}
We note here that in the paper of Chung, Lu and Vu\cite{CLV}
they used the notation $\tilde{d}$ for the second order average degree,
instead of our notation $\langle k \rangle_{2}$ for it.
Applying the above theorem to the AB-model of $\gamma = 3$,
we find that the AB-model belongs to the first category.
Hence, the theorem explains the numerical results\cite{Barabasi,AB,FDBV,GKK,DGMS}.

In spite of such efforts, whether or not the growth of the
maximum eigenvalue of a physical model on the SFN is not so well-known.
This is because the eigenvalues of the adjacency matrix is
different from those of the eigenequation of a physical system.
In this sense, the problem to investigate the growth behavior
of the maximum eigenvalue of the eigenequation of a physical system 
is a nontrivial problem.
So, in order to see this point, 
let us consider the maximum eigenvalue $\omega_{max}$ 
of our vibrational system of oscillators.

We have performed calculations of the
maximum eigenvalue (i.e., vibrational mode) $\omega_{max}$ in our
model of oscillators on the AB-SFN,
where $m = 4$ and $N$ is developed up to $N = 10^{4}$.
This is shown in Fig.2.
The maximum eigenvalue $\omega_{max}^{2}$ (circles),
the maximum degree of a node $k_{max}$ (triangles),
and
the second order average degree $\langle k \rangle_{2}$ of a node (+)
are shown, respectively.
Here we have obtained the following relation:
$$\omega_{0}^{2} + 2\langle k \rangle_{2} \le \omega_{max}^{2} \le \omega_{0}^{2} + 2k_{max}.   \eqno{(21)}$$
This looks similar to the result of Eq.(19) such that
$$\omega_{max}^{2} \propto \omega_{0}^{2} + \sqrt{k_{max}}.    \eqno{(22)}$$
However, this is not supported by our numerical calculations.
Therefore, as the spectral tail of our vibrational model
is different from that of the AB-model as discussed in the previous section,
so is the growth behavior of the maximum eigenvalue of our vibrational model.
This is an important character of our physical model with the AB-SFN.

%%% Figure 2 %%%
\begin{figure}[h]
\includegraphics[scale=0.6]{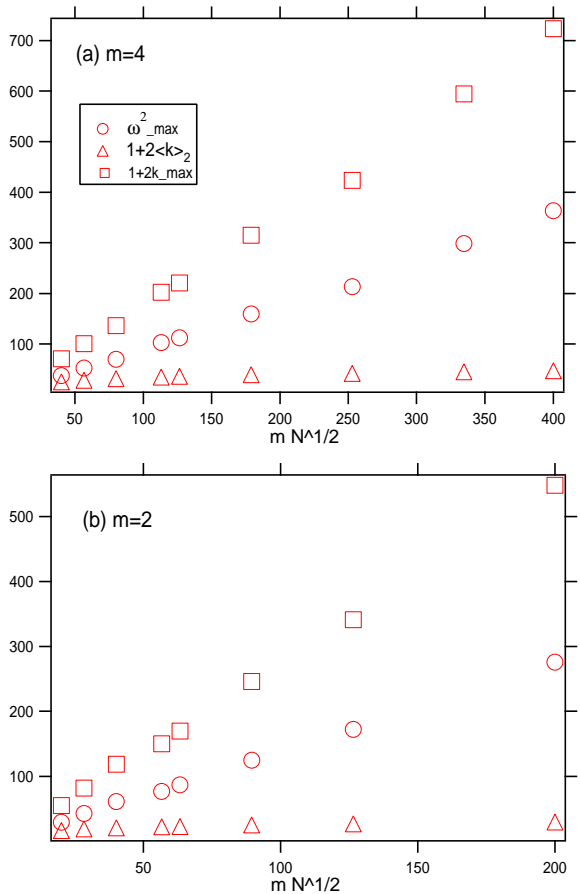}
%\begin{figure}
%\vglue 3.5in
%\hspace{0.1in}
%\special{picture Fig.2 [scaled 0.5]}
%\vglue 0.1in
\caption{
(color online)  
The behavior of the maximum eigenvalue in the spectrum.
The maximum eigenvalue $\omega_{max}^{2}$ (circles),
the maximum degree of nodes $\omega_{0}^{2}+2\langle k \rangle_{2}$ (triangles),
and
the second order average degree $\omega_{0}^{2}+2k_{max}$ of nodes (squares)
are shown in the vertical axis, respectively.
The horizontal axis is scaled as $m\sqrt{N(\tau)}$.
The calculations have been carried out for (a) the 
AB-model with $m = 4$ and $N$ is up to $N = 10^{4}$, and
for (b) the AB-model with $m = 2$ and $N$ is up to $N = 10^{4}$.
Here we have taken the values of $K_{0} = M_{0} = \omega_{0} =1$
and $\langle k \rangle_{2}$ stands for 
the second order average degree of a node.
}
\label{fig:2}
\end{figure}
%%%%%

\section{some theorems on the maximum eigenvalue}
%Hadamard-Gerschgorin theorem
Let us consider the origin of the inequality, Eq.(21).
To see this, let us go back to Eq.(4).
From the Hadamard-Gerschgorin's theorem\cite{Ziman} (see also Appendix A)
we can derive an inequality
$$|\Omega_{i}| \le  K_{0}\sum_{j=1}^{N(\tau)} |A_{ji}|\frac{|q_{j}|}{|q_{i}|}. \eqno{(23)}$$
Since $|q_{j}|/|q_{i}| \le 1$ and $|A_{ji}| = A_{ji}$, we can derive
$|\Omega_{i}| \le K_{0}\sum_{j}|A_{ji}| = K_{0}k_{i}(\tau)$,
which then yields a theorem:
\begin{theorem}
$$\left|\omega^{2}-\omega_{i}^{2} - \frac{K_{0}}{M_{i}}k_{i}(\tau) \right| 
\le \frac{K_{0}}{M_{i}}\sum_{j}|A_{ji}| = \frac{K_{0}}{M_{i}}k_{i}(\tau).\eqno{(24)}$$
\end{theorem}
Thus, there exists at least one atomic site (i.e., node) that
satisfies Eq.(24) for all eigenvalues $\omega$.
This implies that $\omega^{2}$ is included within a disk of radius 
$\frac{K_{0}}{M_{i}}k_{i}(\tau)$
and its center $\omega_{i}^{2} + \frac{K_{0}}{M_{i}}k_{i}(\tau)$.

%Perron-Frobenius's theorem
Since in our model all $K_{ij}$ (i.e., $A_{ij}$) are non-negative,
by applying the Perron-Frobenius's theorem\cite{Ziman} (see also Appendix B)
to Eq.(24) we can derive 
$|\omega^{2}-\omega_{i}^{2}| \le  2\frac{K_{0}}{M_{0}}k_{i}(\tau)$.
Hence, the maximum frequency $\omega_{max}^{2}$
satisfies another theorem:
\begin{theorem}
$$\omega_{max}^{2}  \le 
\max_{i}\left[\omega_{i}^{2} + 2\frac{K_{0}}{M_{i}}k_{i}(\tau)\right].           \eqno{(25)}$$
\end{theorem}
And similarly we can obtain a more precise theorem:
\begin{theorem}
$$\min_{i}\left[\omega_{i}^{2} + 2\frac{K_{0}}{M_{i}}k_{i}(\tau)\right] 
\le \omega_{max}^{2} 
\le \max_{i}\left[\omega_{i}^{2} + 2\frac{K_{0}}{M_{i}}k_{i}(\tau)\right].       \eqno{(26)}$$
\end{theorem}
From this, if we assume that $M_{i}\equiv M_{0} =$ const. 
and $\omega_{i} = \omega_{0} =$ const.
and applying for the SFN, then Eq.(25) becomes 
\begin{theorem}
$$2\frac{K_{0}}{M_{0}}\min_{i}\{k_{i}(\tau)\} 
\le \omega_{max}^{2} - \omega_{0}^{2}
\le 2\frac{K_{0}}{M_{0}}\max_{i}\{k_{i}(\tau)\}.                     \eqno{(27)}$$
\end{theorem}
Hence, this theorem verifies our numerical results 
in the previous section.
Therefore, the upper limit of the spectrum (i.e., spectral edge) 
grows as fast as the network grows.
This is a remarkable fact for excitations in the SFN models
and this nature is very different from that of Anderson localization
where only mobility edge may appear in the spectrum and 
the band edge cannot grow as fast as the system size grows\cite{Ander,Ziman}.

The above Theorems 2-5 are good for the standard networks
that the distribution of the orders of nodes
is limited such as periodic lattice systems or the RN\cite{Ziman}
or
the SWM\cite{Klein,WS,NMWS,CNSW,NSW,BA,ASBS,LCS,Strog},
since in these systems there exist finite lower and upper limits of the orders of nodes
such that the error width is bounded as 
$$\Delta(\omega_{max}^{2}-\omega_{0}^{2}) \ge 
\max_{i}[2\frac{K_{0}}{M_{i}}k_{i}(\tau)]
-\min_{i}[2\frac{K_{0}}{M_{i}}k_{i}(\tau)].                  \eqno{(28)}$$
However, whether or not the above theorems can be accurate 
conditions for the SFN\cite{Barabasi,AB} is not trivial,
since in the SFNs there exist various orders of nodes without any bound but 
with the power-law distribution.

To study this point, we first observe that 
there is a particularly important nature of the
adjacency matrix ${\bf \hat{A}}$ in the network theory.
Denote by 
$\vec{k}_{i} = (A_{i1},\dots , A_{iN(\tau)})^{t}$
the $i$-th column vector of ${\bf \hat{A}}$.
The vector represents the way of links between 
the $i$-th node and other linked nodes,
such that it defines the order $k_{i}$ 
of the $i$-th node such that
$$k_{i} \equiv \vec{k}_{i}^{t}\cdot \vec{k}_{i} = \sum_{j}A_{ij}.  \eqno{(29)}$$
Therefore, let us call $\vec{k}_{i}$ vectors the {\it link vectors}.
Using this representation, we can rewrite the adjacency matrix as
${\bf \hat{A}} = (\vec{k}_{1}, \dots, \vec{k}_{N(\tau)})
= (\vec{k}_{1}^{t}, \dots, \vec{k}_{N(\tau)}^{t})^{t}$,
where $t$ means the transpose.
From this, we can derive that
$${\bf \hat{A}}^{2} = (\vec{k}_{i}^{t}\cdot \vec{k}_{j}),    \eqno{(30)}$$
which is a symmetric matrix and
nothing but the Gramian matrix between 
the link vectors, $\vec{k}_{i}$, where
$$Tr({\bf \hat{A}}^{2}) = \sum_{i=1}^{N(\tau)}\vec{k}_{i}^{t}\cdot \vec{k}_{i}
= \sum_{i=1}^{N(\tau)}k_{i} = 2L(\tau).                      \eqno{(31)}$$

Let us go back to Eq.(4).
We now rewrite it as
$\Omega_{i}\vec{q} = K_{0}{\bf \hat{A}}\vec{q}$.
Therefore, $\Omega_{i}^{2}\vec{q} = K_{0}^{2}{\bf \hat{A}}^{2}\vec{q}$.
Let us now use the Hadamard-Gerschgorin theorem\cite{Ziman} (Appendix A)
or the Perron-Frobenius theorem\cite{Ziman} (Appendix B) for 
$K_{0}^{2}{\bf \hat{A}}^{2}$,
we can derive an inequality
$$|\Omega_{i}^{2}| \le  K_{0}^{2}\sum_{j=1}^{N(\tau)} 
|({\bf \hat{A}}^{2})_{ji}|\frac{|q_{j}|}{|q_{i}|}.                \eqno{(32)}$$
Since $|q_{j}|/|q_{i}| \le 1$ and $|({\bf \hat{A}}^{2})_{ji}| 
= ({\bf \hat{A}}^{2})_{ji} = \vec{k}_{j}^{t}\cdot \vec{k}_{i}$, 
we can derive
$|\Omega_{i}^{2}| \le K_{0}^{2}\sum_{j}|({\bf \hat{A}}^{2})_{ji}| 
= K_{0}^{2}\sum_{j}\vec{k}_{j}^{t}\cdot \vec{k}_{i}
= K_{0}^{2}\vec{k}_{tot}^{t}\cdot \vec{k}_{i}$,
where
$$\vec{k}_{tot}^{t} = \sum_{i=1}^{N(\tau)}\vec{k}_{i}^{t}
= (k_{1}, k_{2}, \dots, k_{N}).        \eqno{(33)}$$
Then we have
$$\left| \omega^{2}-\omega_{i}^{2} - \frac{K_{0}}{M_{i}}k_{i}(\tau)\right|^{2} 
\le \left(\frac{K_{0}}{M_{i}}\right)^{2}\vec{k}_{tot}^{t}\cdot \vec{k}_{i}.     \eqno{(34)}$$
Therefore, it then yields a theorem:
\begin{theorem}
$$\left| \omega^{2}-\omega_{i}^{2} - \frac{K_{0}}{M_{i}}k_{i}(\tau)\right|
\le \frac{K_{0}}{M_{i}} \sqrt{\vec{k}_{tot}^{t}\cdot \vec{k}_{i}}.     \eqno{(35)}$$
\end{theorem}
Thus, there exists at least one atomic site (i.e., node) that
satisfies Eq.(35) for all eigenvalues $\omega$.
This implies that $\omega^{2}$ is included within a disk of radius 
$\frac{K_{0}}{M_{i}}\sqrt{\vec{k}_{tot}^{t}\cdot \vec{k}_{i}}$
and its center $\omega_{i}^{2} + \frac{K_{0}}{M_{i}}k_{i}(\tau)$.
Since
$\left| \omega^{2}-\omega_{i}^{2}\right| - \left|\frac{K_{0}}{M_{i}}k_{i}(\tau)\right|
\le \left| \omega^{2}-\omega_{i}^{2} - \frac{K_{0}}{M_{i}}k_{i}(\tau)\right|$,
we obtain
$$\left| \omega^{2}-\omega_{i}^{2}\right| \le 
\frac{K_{0}}{M_{i}}\left(k_{i}(\tau) 
+\sqrt{\vec{k}_{tot}^{t}\cdot \vec{k}_{i}}\right). \eqno{(36)}$$
Therefore, for the maximum frequency
we obtain
$$\left| \omega_{max}^{2}-\omega_{i}^{2}\right| \le 
\max_{i}\left[\frac{K_{0}}{M_{i}}\left(k_{i}(\tau) 
+\sqrt{\vec{k}_{tot}^{t}\cdot \vec{k}_{i}}\right)\right]. \eqno{(37)}$$
Hence, we obtain
$$\omega_{max}^{2} \le \omega_{i}^{2} + 
\max_{i}\left[\frac{K_{0}}{M_{i}}\left(k_{i}(\tau) 
+\sqrt{\vec{k}_{tot}^{t}\cdot \vec{k}_{i}}\right)\right]. \eqno{(37)}$$
Since
$k_{i} \le \vec{k}_{tot}^{t}\cdot \vec{k}_{i}$,
we have
$\max_{i}\left[\frac{K_{0}}{M_{i}}\left(k_{i}(\tau) 
+\sqrt{k_{i}(\tau)}\right)\right] \le 
\max_{i}\left[\frac{K_{0}}{M_{i}}\left(k_{i}(\tau) 
+\sqrt{\vec{k}_{tot}^{t}\cdot \vec{k}_{i}}\right)\right]$.
Therefore, the right hand of Eq.(37) is comparable
with that of Eq.(26).
In this way, Theorems 2-5 work for the SFN systems as well.

\section{conclusions}
In conclusion, we have studied the system of oscillators
connected by springs in the geometry of the AB SFN model.
We first presented the Green function formalism for obtaining
the spectrum of the vibrational modes of the system. 
In the case of very weak spring constant,
using this formalism
we find that the distribution of eigenmodes follows 
the same type of power-law distribution
of degrees of a node in the AB model [see Eq.(13)]
such that 
$P(\omega^{2}-\omega_{0}^{2}) \propto (\omega^{2}-\omega_{0}^{2})^{-\gamma}$
with $\gamma = 3$.
In the case of an arbitrary strength of spring constants,
we have performed numerical calculations in order to
obtain the spectrum of vibrational modes.
We have found that even in this case,
the distribution of eigenmodes obeys the same type of the
power-law distribution of degrees of a node in the AB model as well
[see Fig.1].
This is contrary to the distribution of eigenvalues 
of adjacency matrix in the AB model,
where power-law distribution is given by 
$\rho(\lambda) \propto \lambda^{-\gamma'}$
with $\gamma' = 2\gamma -1 = 5$.

This is a consequence of our model, where
relative displacements between the
individual oscillators are included in the Hamiltonian.
This Hamiltonian provides the diagonal matrix elements in the
eigenequation, which
are proportional to the degrees of nodes [see Eq.(4)].
These diagonal elements can be regarded as on-site potentials
in the problem.  
Since the degree of a node develops indefinitely,
the on-site potential can be arbitrary large as
the system is progressing.
Therefore, the eigenvalues are strongly dominated by the
magnitude of the diagonal elements of the eigenequation.
Thus, the distribution of eigenmodes is affected by
that of degrees of nodes such that
the distribution of eigenmodes coincides with that
of degrees of nodes in the network.

We finally have investigated the asymptotic behavior
of the maximum eigenvalue $\omega_{max}$ of the system.
We have found numerically that the maximum eigenvalue
is bounded as in Eq.(21).
From this, as the total number of nodes, $N$, 
is increasing, 
the maximum degree of nodes becomes arbitrarily large.
Therefore, the maximum eigenvalue can be arbitrarily
large as $N \rightarrow \infty$.
This coincides with the result of the maximum
eigenvalue of adjacency matrix in the AB model.
We have also proved the above numerical results
by some mathematical theorems that are proved
using the Hadamard-Gerschgorin theorem 
and the Perron-Frobenius theorem.

Thus we conclude that when we apply
a certain physical model to the geometry of a SFN,
the physical properties are strongly dominated by the
nature of the SFN.
In this sense, not only the network geometry of a SFN
but also the property of physical models on a SFN
are important in the study of the SFN.
This direction will be very interesting
for further researches.

\acknowledgments
This paper is dedicated to the memory of Dr. Mihoko Yoshida
(Lehigh University)
who always helped us for collecting relevant papers.
We would like to thank Dr. Jun Hidaka for sending us
relevant papers.
K. I. would like to thank Kazuko Iguchi for her
financial support and encouragement.

\appendix
%\secction{appendix A: the Hadamard-Gerschgorin Theorem}
\section{The Hadamard-Gerschgorin Theorem}
The following theorem is known as 
the  Hadamard-Gerschgorin theorem in linear algebra\cite{Ziman}.
Consider the following eigenequation:
$$(\lambda - h_{i})q_{i} = \sum_{j = 1 (\ne i)}^{N}h_{ij}q_{j},\eqno{(A.1)}$$ 
for $i = 1, \dots, N$.
Then, we find
$$|\lambda - h_{i}| \le \sum_{j = 1 (\ne i)}^{N}|h_{ij}|\frac{|q_{j}|}{|q_{i}|},\eqno{(A.2)}$$
for $i = 1, \dots, N$.
Since always $\frac{|q_{j}|}{|q_{i}|} \le 1$, we obtain
$$|\lambda - h_{i}| \le \sum_{j = 1 (\ne i)}^{N}|h_{ij}| \equiv B_{i}, \eqno{(A.3)}$$
for $i = 1, \dots, N$.
Now, we find a theorem
that there exists at least one site such that the above equation Eq.(A.3)
is valid for all $\lambda$.
Eq.(A.3) means that $\lambda$ is included within 
a disk of radius $B_{i}$ with its center of $h_{i}$.

%\section{appendix B: the Perron-Frobenius Theorem}
\section{The Perron-Frobenius Theorem}
The following theorem is known as 
the Perron-Frobenius theorem in linear algebra\cite{Ziman}.
Suppose that an $n\times n$ symmetric matrix $H$
has all non-negative entries $h_{ij} \ge 0$.
Then this satisfies an eigenequation
$H|\psi_{i} \rangle = \lambda_{i} |\psi_{i}\rangle$.
For any positive constants $c_{1}, c_{2}, \dots, c_{n}$,
the maximum eigenvalue $\lambda_{max}(H)$ satisfies
$$\lambda_{max}(H) \le \max_{1 \le i \le n} 
\left\{ \sum_{j=1}^{n} \frac{c_{j} h_{ij}}{c_{i}}\right\}. $$

%\end{multicols}
\end{document}